\documentclass[10pt]{article} 
\usepackage{spie}
\usepackage{latexsym}
\usepackage{times,mathptm} 
\usepackage{graphicx} 
\usepackage{multicol}
\usepackage{rotating} 
\usepackage[verbose]{wrapfig}
\usepackage{epsfig}
\newlength{\dinwidth}
\newlength{\dinmargin}
\newcommand{\bc}{\begin{center}}
\newcommand{\ec}{\end{center}}
\newcommand{\be}{\begin{equation}}
\newcommand{\ee}{\end{equation}}
\newcommand{\bi}{\begin{itemize}}
\newcommand{\ei}{\end{itemize}}
\newcommand{\bt}{\begin{table}}
\newcommand{\enta}{\end{table}}

\newcommand{\g}{$\gamma$}

\newcommand{\na}{$^{22}$Na\ }

\newcommand{\yt}{$^{88}$Y\ }

\newcommand{\beq}{\begin{equation}}
\newcommand{\eeq}{\end{equation}}
\newcommand{\bfg}{\begin{figure}}
\newcommand{\efg}{\end{figure}}
\newcommand{\keV}{\mbox{ke\hspace{-0.1em}V}}
\newcommand{\MeV}{\mbox{Me\hspace{-0.1em}V}}

\newcommand{\cphibar}{\ensuremath{\cos{\bar{\varphi}}}}
\newcommand{\cphigeo}{\ensuremath{\cos{\varphi^\triangleleft}}}
\newcommand{\ephibar}{\ensuremath{\sigma_{\cos{\bar{\varphi}}}}}
\newcommand{\ephigeo}{\ensuremath{\sigma_{\cos{\varphi^\triangleleft}}}}
\newcommand{\phibar}{\ensuremath{\bar{\varphi}}}
\newcommand{\phigeo}{\ensuremath{\varphi^\triangleleft}}
\renewcommand{\deg}{\ensuremath{^\circ}}
\newcommand{\gcc}{\mbox{g cm$^{-3}$}}
\graphicspath{{figures/}}
\begin{document}
\title{Compton scattering sequence reconstruction algorithm for the liquid xenon
gamma-ray imaging telescope (LXeGRIT)} 
\author{U.G. Oberlack, E. Aprile, A. Curioni, V. Egorov, K.L. Giboni \\ 
Columbia Astrophysics Laboratory, Columbia University, New York, USA} 
\authorinfo{Send correspondence to: U.G. Oberlack, Columbia University, 
Astrophysics Laboratory, 550 West 120th Street, New York, NY 10027 \\
E-mail: oberlack@astro.columbia.edu \hfill \\
LXeGRIT Web page: \texttt{http://www.astro.columbia.edu/$\sim$lxe/lxegrit/}
}
\pagestyle{plain}    

\maketitle

\begin{abstract} 

The Liquid Xenon Gamma-Ray Imaging Telescope (LXeGRIT) is a balloon born
experiment sensitive to \g --rays in the energy band of 0.2--20~\MeV. The main
detector is a time projection chamber filled with high purity liquid xenon
(LXeTPC), in which the three--dimensional location and energy deposit of individual \g
--ray interactions are accurately measured in one homogeneous volume. 
To determine the \g --ray initial direction (Compton imaging), as well as to
reject background, the correct sequence of interactions has to be determined.  
Here we report the development and optimization of an algorithm to reconstruct
the Compton scattering sequence and show its performance on Monte Carlo events
and LXeGRIT data.

\end{abstract}


\section{Introduction}
\begin{sloppypar}
\begin{wrapfigure}[19]{R}{0.42\textwidth}
\centering
\includegraphics[width=0.5\linewidth, clip]{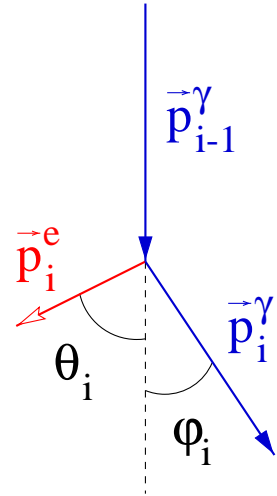}
\caption{\label{f:kinematics} Kinematics of Compton scattering.}  
\end{wrapfigure}
In order to overcome the relatively low detection efficiency ($< 1\%$)
intrinsic of a double scatter Compton telescope such as COMPTEL \cite{VSchon:93} 
a homogeneous, self--triggered, three--dimensional (3D) position sensitive
detector such as a LXeTPC,
was proposed several years ago \cite{EAprile:89:IEEE}~.
In a COMPTEL type telescope the acceptable event topologies are restricted
to a single Compton scatter in a first detector layer ({\it
converter}), followed by absorption of the scattered \g --ray in a second
detector layer ({\it absorber}), with the separation between the two layers
large 
enough to order the two interactions by time-of-flight (TOF) measurement. In a
homogeneous detector, however,  detection of \MeV \ \g --rays allows a
variety of different event topologies to be recorded and used for imaging, with
a large increase in efficiency.
In absence of a TOF measurement, the correct order of the multiple Compton
interactions has to be determined from the redundant kinematical and geometrical
information measured for each event.
As previously suggested in Aprile et al. 1993 \cite{EAprile:93:NIM} and more
recently in Schmid et al. 1999 \cite{GJSchmid:99:tracking} and in Boggs et
al. 2000 \cite{SBoggs:00:comp_reconst}~, 
event reconstruction based on Compton kinematic allows to correctly order the
interactions. Moreover, background suppression is a direct consequence 
Compton kinematic reconstruction. Here we present the current status of
our work on this topic within the context of the LXeGRIT program.

\section{Definition of the problem}

The kinematic of Compton scattering is displayed in Fig.~\ref{f:kinematics}. 
Energy and momentum conservation in the Compton scatter process are written as
\begin{eqnarray}
  E_{i-1}^{\gamma} &=& E_{i}^{\gamma} + E_i^\mathrm{e} \label{e:energy} \\
  \vec{p}_{i-1}^{\gamma} &=& \vec{p}_{i}^{\gamma} + \vec{p}_i^\mathrm{e}
  \qquad (i=1, \ldots, N-1)
  \label{e:momentum}
\end{eqnarray}
with $E_{i}^{\gamma} \; (i=0 ,\ldots, N-1)$ and $E_i^\mathrm{e}\; (i=1 ,\ldots,
N)$ the energy of the \g-photon and the scattered electron after interaction
$i$. $E_{0}^{\gamma}$ is the energy of the incoming photon and $\vec{p}_{i}$ are
the corresponding momenta.
The dispersion relations for photons and electrons reduce these four equations
to three independent equations. The vector equation~\ref{e:momentum} translates
into two independent equations for the photon scatter angle $\varphi_i$ and the
electron scatter angle $\theta_i \; (i=1, \ldots, N-1)$: 
\begin{eqnarray}
  \cos \varphi_i  &=& 1 - \frac{1}{W_i} + \frac{1}{W_{i+1}} \textrm{, with:}
  \quad W_i = \frac{E_i}{m_0 c^2} 
  \label{e:phi} \\
  \cot \theta_i &=& \left( 1+\frac{E_{i-1}^{\gamma}}{m_0 c^2} \right) 
                 \: \tan\frac{\varphi_i}{2}  \label{e:theta} 
\end{eqnarray}
If the last measured interaction $N$ is a photo-absorption, there is no
meaningful electron scatter angle $\theta_N$ (even though an electron is
released in the process). 
For LXeGRIT, the electron scatter angles are not measured (in the considered
energy range, interactions are actually pointlike, since the typical electron
range does not exceed the given granularity) and are therefore ignored in the
following.  
The detector measures $N$ energy deposits $E_i (\approx E_i^\mathrm{e})$ and $N$
interaction locations $\vec{x}_i$. For a given interaction sequence, the
locations determine geometrically $N-2$ photon scatter angles \phigeo$_i \
(i=2,\ldots, N-1)$ in the usual case of unknown source position: 
\be
\cphigeo_i = \frac{\overrightarrow{u} _i \cdot \overrightarrow{u}
_{i+1}}{|\overrightarrow{u} _i | |\overrightarrow{u}_{i+1} |}
\ee
where $\overrightarrow{u} _i ~=~ (x_i-x_{i-1},\: y_i-y_{i-1},
\:z_i-z_{i-1})$. \\
For calibration sources, \phigeo $_1$ is known additionally. 
Moreover, $N-1$ Compton
scatter angles \phibar$_i$ are measured by the energy deposits according to
equation~\ref{e:phi}, noting that $E_i^{\gamma} = \sum_{j=i+1}^{N} E_j \; (i =
0,\ldots,N-1)$. This redundant information allows testing of the sequence of the
interaction points based solely on kinematics. A straightforward test statistic
consists of summing the differences of the scatter angles quadratically:
\begin{equation} \label{e:testa:prime}
  T_\varphi^\prime = \sum_{i=2}^{N-1} (\cos{\phibar_i} - \cos{\phigeo_i})^2
\end{equation}
Ideally, the test statistic would be zero for the correct sequence if the photon
is fully contained. With measurement errors, $T_\varphi^\prime$ is always
greater than zero, but the correct interaction sequence is still most likely to
produce the minimum value of the test statistic. In addition to minimizing $T$,
an upper threshold can in principle discriminate against photons that are not
fully absorbed. 
This can be improved by weighting the summands with the measurement errors:
\begin{eqnarray} \label{e:testa}
  T_\varphi &=& \sum_{i=2}^{N-1} \frac{(\cos{\phibar_i} -
  \cos{\phigeo_i})^2}{\sigma_i^2} \\ 
  \textrm{with:} \quad \sigma_i^2 &=& \sigma_{\cos{\phibar},i}^2 +
  \sigma_{\cos{\phigeo},i}^2 
  \nonumber 
\end{eqnarray}
For each triplet of interactions \ephibar \ and \ephigeo \ are computed
in the following way:   
\begin{eqnarray} \label{e:errors}
\sigma_{\cos{\phigeo},i}^2 &=&
\frac{2}{|\overrightarrow{u} _i| ^2 \cdot |\overrightarrow{u} _{i+1}| ^2}
\sum_{k=1}^{3} (2u_{i, k} ^2 +2u_{i+1, k} ^2 - 2u_{i, k} u_{i+1, k})\cdot
\sigma _k ^2 \\ 
\textrm{with:} &k& \ \textrm{spatial coordinate index and} \nonumber \\ 
&\sigma _k & \ \textrm{position uncertainty on each coordinate} \nonumber \\ 
\sigma_{\cos{\phibar},i}^2 &=&
\frac{1}{W_i ^4}\cdot \sigma (W_i - W_{i+1}) ^2 + \left
( \frac{1}{W_i ^2} - \frac{1}{W_{i+1} ^2} \right) ^2 \cdot \sigma (W_{i+1}) ^2
\end{eqnarray}
Eq. \ref{e:errors} holds only under assumption that the 3D separation
between two interactions is large compared to position uncertainties, which we
insure by corresponging data cuts.	
\end{sloppypar}

\section{The LXeGRIT Detector}

\begin{figure}[htb]
\begin{center}
\psfig{file=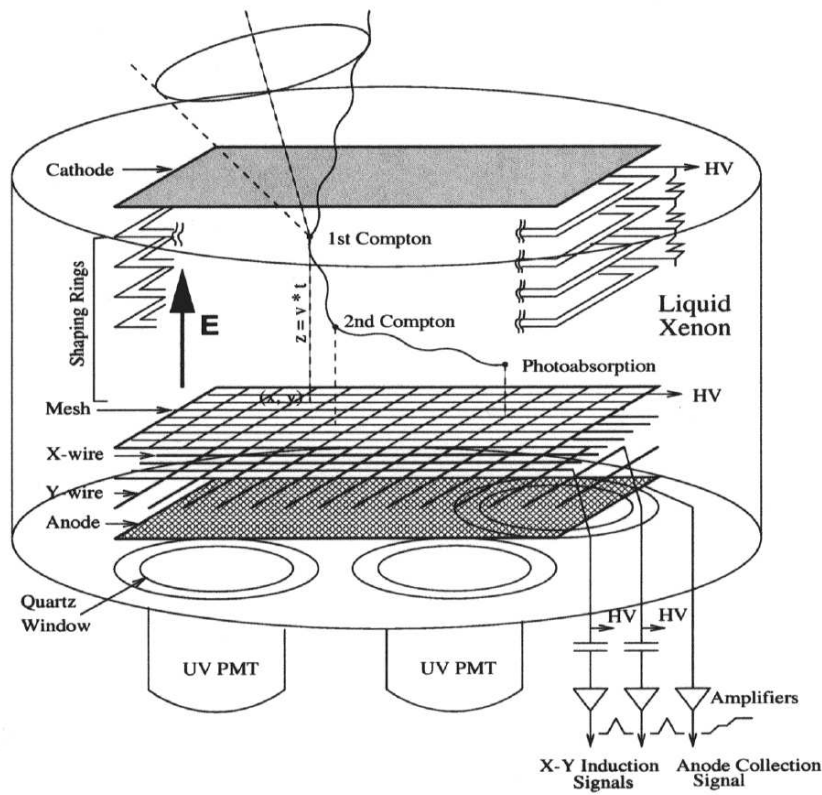,width=0.7\linewidth,clip=}
\caption{\label{f:TPC:schem}LXeTPC schematic.}
\end{center}
\end{figure}

The LXeGRIT (Liquid Xenon Gamma-Ray Imaging Telescope) instrument was
developed as a prototype of Compton telescope exploiting 3D
imaging capability and good spectral response of a LXeTPC. 
It is described in Aprile et al. \cite{EAprile:98:electronics}$^,$
\cite{EAprile:2000:pisa2000} .  
The principle of operation of the LXeTPC is schematically shown in Fig. \ref{f:TPC:schem}.
The TPC is assembled in a cylindrical vessel of 10~l 
volume, filled with high purity liquid xenon (LXe). The sensitive area is $20
\times 20$~cm$^2$ and the drift region is 7 cm. The 
detector  operates over a wide energy range from $\sim$~200~\keV\ to 20~\MeV.  
Both ionization and scintillation light signals are detected. Ionization
signals measure energy and 3D position for each interaction, while the fast
($<5$~ns) Xe scintillation light provides the event trigger.  
The drift of free ionization electrons in the uniform  electric field,
typically 1~kV/cm, induces charge signals on a pair of orthogonal planes of
parallel wires with a 3~mm pitch, before collection on four independent
anodes. Each of the 62 X-wires and 62 Y-wires and each anode is amplified and
digitized at a sampling rate of 5~MHz, to preserve the pulse shape. The X-Y
coordinate information is obtained from the pattern of hits on the wires, while
the energy is obtained from the amplitude of the anode signals. The
Z-coordinate is determined from the drift time measurement, referred to the
light trigger.
Relying on low-noise readout electronics (less than $\sim$~400~e$^-$~RMS on the
wires, $\sim$~800~e$^-$~RMS on the anodes), the TPC can well detect the multiple
interactions of \MeV \ \g--rays, with a minimum energy deposit as low as
100~\keV. \\
A typical event is displayed in Fig. \ref{f:sig_rec}. The
digitized signals on all wires and active anodes are shown as a function of 
drift time. The incoming photon makes two Compton scatterings befor being
photoabsorbed. In this 3-interaction event, in fact, the sum of the three anode
pulse heights gives 1.8~\MeV. The corresponding locations of the interactions are
clearly seen on the X-Y wires. \\  
The energy resolution, at 1~kV/cm electric field, has been measured to be $\sim
9\%$ FWHM at 1~\MeV, scaling as $1/\sqrt{E}$: the position resolution is $\sim
1$~mm on the X-Y coordinates, as expected for the given granularity (3~mm spaced
wires), and 0.15~mm on the Z coordinate, determined by the accuracy on the drift
time measurement. \\ 
Once digitized, wire and anode signals are passed through a signal
recognition and fitting algorithm, which sorts the 
different event topologies and reduces the event to the relevant information,
i.e. X-Y-Z location and energy for each interaction in the sensitive volume. \\ 
\bfg[htb]
\begin{center}
\psfig{file=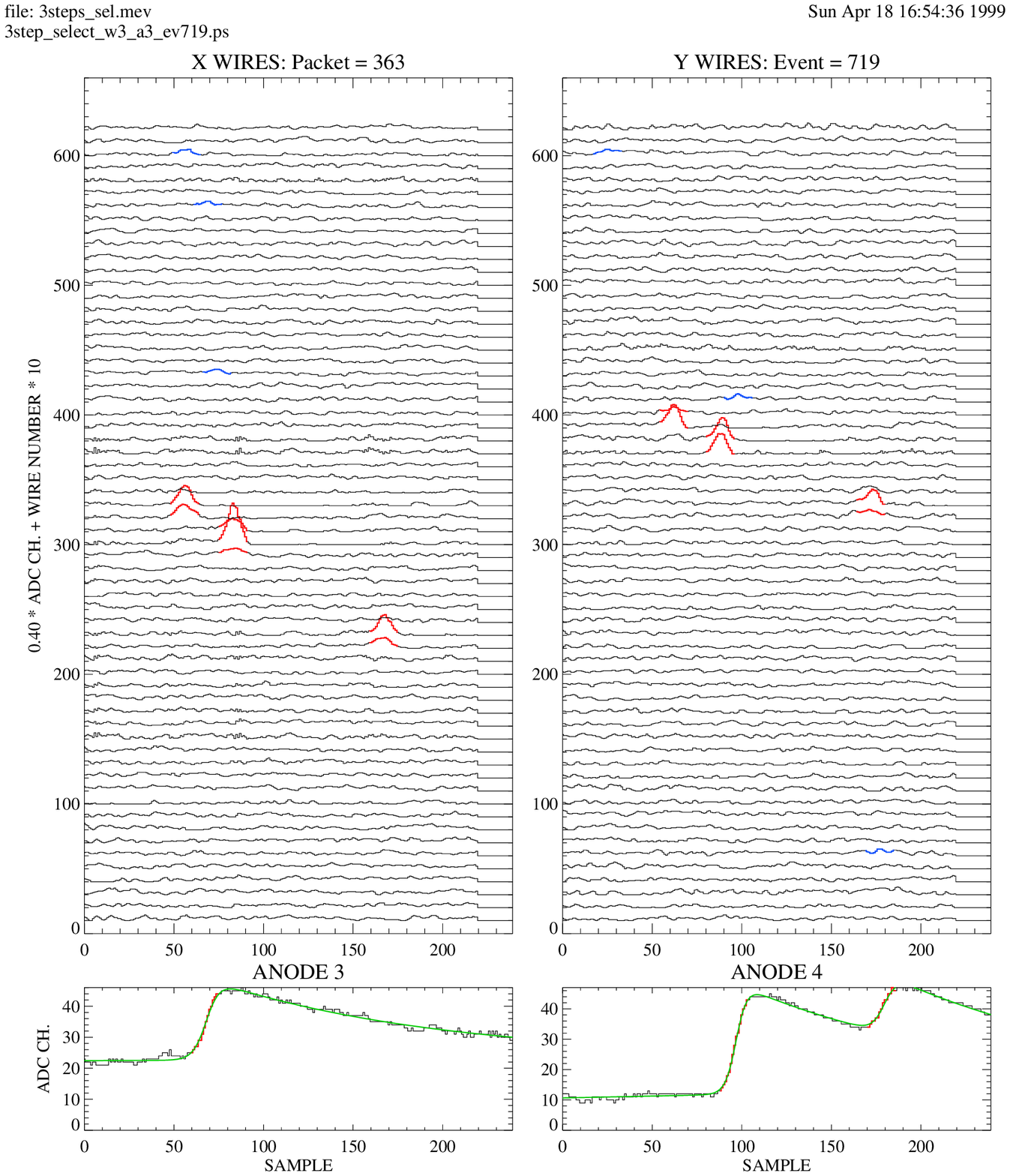,bbllx=62,bblly=143,bburx=563,bbury=726,%
       width=.65\linewidth,clip=}
\end{center}
\caption{\label{f:sig_rec} On-line display of a 3-interaction event (two Compton
scatters followed by photoabsorption).}  
\efg
\section{Performance of the algorithm}

The Compton sequence reconstruction (CSR) algorithm has been implemented in a C
routine, as a module of the LXeGRIT analysis software. It has been tested using
Monte Carlo data and then applied to experimental data. \\ 
Given the range of tipically recognizable interactions, $3~\leq$~N~$\geq~6$, the
algorithm currently performs a ``brute force'' search through all N factorial
(N!) permutations of the sequence. For each sequence
the algorithm first checks whether the sequence is kinematically forbidden
(i.e. \cphibar $_i$ ~$<~-1$ ), in which case the sequence is rejected. Otherwise it
calculates the test statistic (defined in eq. \ref{e:testa}).
For each event the sequence which minimizes the test statistic is then selected
as the true Compton scattering sequence. \\
Rejection of non-Compton events or discrimination between
fully and partially absorbed photons, based on some upper threshold on the
minimum value of the test statistic, is a desirable feature of the
algorithm. The topic of background rejection will be discussed elsewhere in a
broader contest. In this paper, an upper threshold on the test
statistic has been applied only to reduce the fraction of incorrectly
reconstructed Compton events.

\subsection{\label{s:MC} Monte Carlo Data}

\begin{figure}
\bc
\epsfig{file=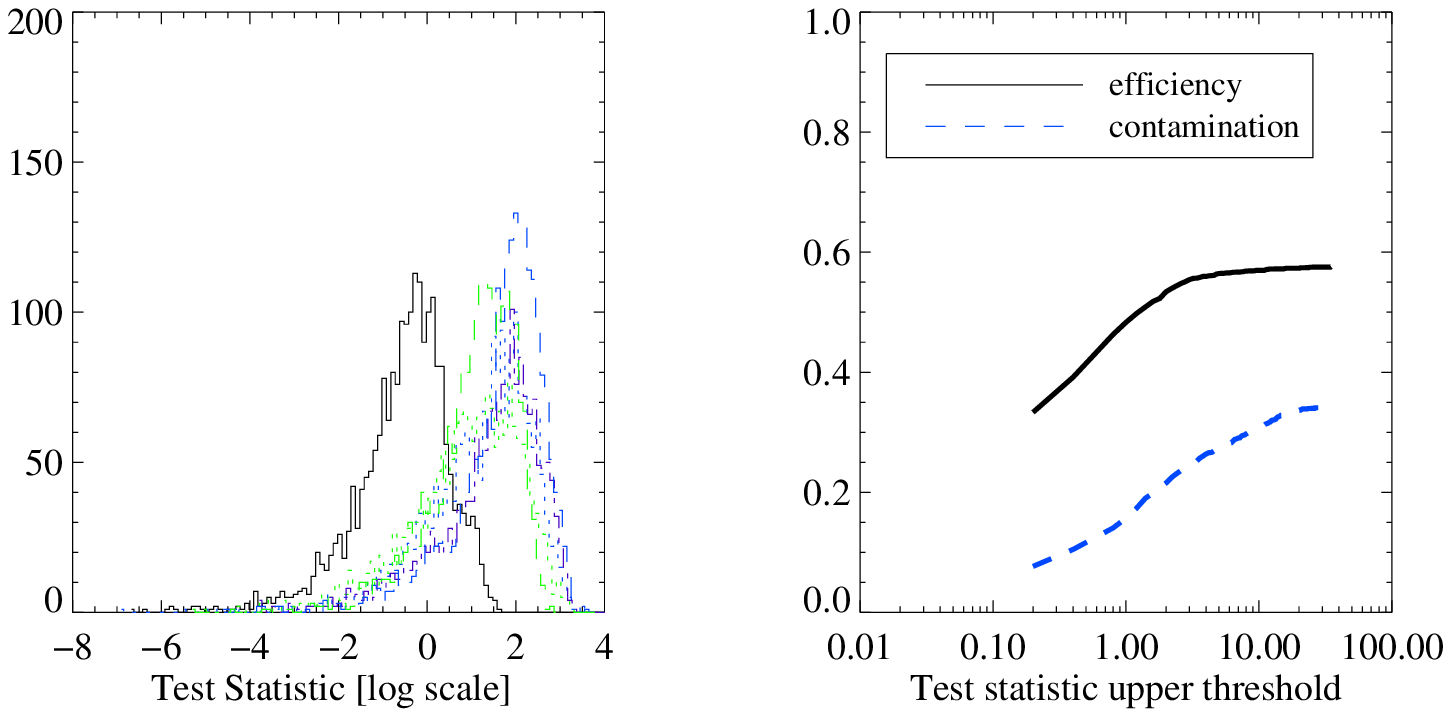,bbllx=100,bblly=530,bburx=540,bbury=750,width=0.9\linewidth,clip=}
\ec
\caption{\label{f:MC1} Performance of the CSR algorithm tested with Monte Carlo
generated events in the 1836~\keV \ 
photopeak. {\it Left:}~the test statistic for the true sequence (continuos
line) and for the remaining five false sequences, {\it right:}~fraction of correctly and
incorrectly reconstructed events ({\it efficiency} and {\it confusion}) as a
function of upper threshold on test statistic.}
\end{figure}

\begin{figure}
\bc
\epsfig{file=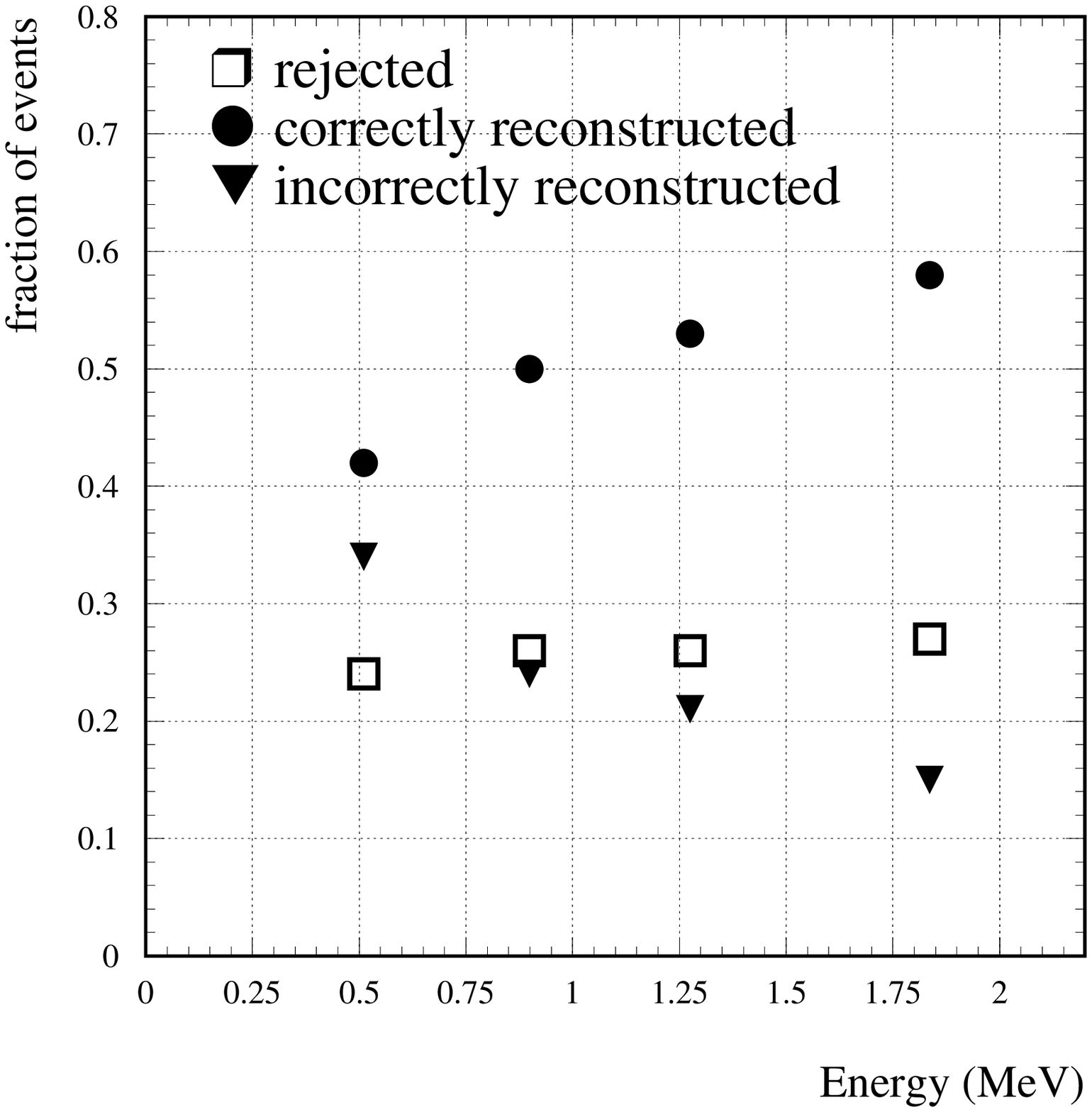,height=9.cm,width=0.7\linewidth,clip=}
\ec
\caption{\label{f:eff} Efficiency, contamination and the fraction of rejected
         events as a function of energy for an upper threshold of 2 on test
         statistic.} 
\end{figure}

The performance of the algorithm on Monte Carlo data is shown in
Figs.~\ref{f:MC1} and \ref{f:eff}. 
The events have been generated using the GEANT detector simulation
package \cite{GEANT}.
\g --rays of different energies have been generated. Here we consider 511~\keV \  and 
1275~\keV, i.e. the two lines from \na\ decay, 898~\keV \  and 1836~\keV, from
\yt (as the two sources have been widely used in calibrating the detector: the
source location during data taking has been reproduced in the simulation).
Each \g --ray
is tracked through a detailed mass model of the LXeGRIT detector. For each
interaction, X-Y-Z location, energy deposit and interaction type are
recorded. The following conditions and experimental uncertainties are then
imposed on the simulated events to reproduce the LXeGRIT events:

\bi
\item 100~\keV \ minimum energy threshold for detection of each interaction 
\item 1~mm RMS position resolution on each coordinate
\item 10$\%$ FWHM at 1~\MeV, scaling as $1/\sqrt{E}$, energy resolution.
\item 5~mm minimum spatial separation on at least one coordinate between each
      pair of interactions in order to consider two interaction locations mutually
      resolved. If this condition is not fullfilled the two interactions
      are clustered and considered as one single interaction 
\ei

We focus on events with three detected interactions (the minimum number required
in the algorithm) since this event class dominates over higher multiplicities in
LXeGRIT. 
Fig. \ref{f:MC1} summarizes the results for events in the 1836~\keV \
full-energy peak. On the left panel test statistic for the true sequence and 
for the remaining five false sequences is shown. The right panel displays the
fraction of correctly and incorrectly reconstructed events ({\it efficiency} and
{\it confusion}) as a function of the upper threshold applied to test statistic. \\ 
Efficiency, contamination and the fraction of rejected events are shown as a
function of energy in Fig. \ref{f:eff} for an upper threshold of 2 on test statistic.
For energies larger than 1~\MeV\ a reconstruction efficiency larger than 50$\%$
is achieved.   

%
%

\subsection{Experimental Data}
\bfg[htp]
\begin{center}
\psfig{file=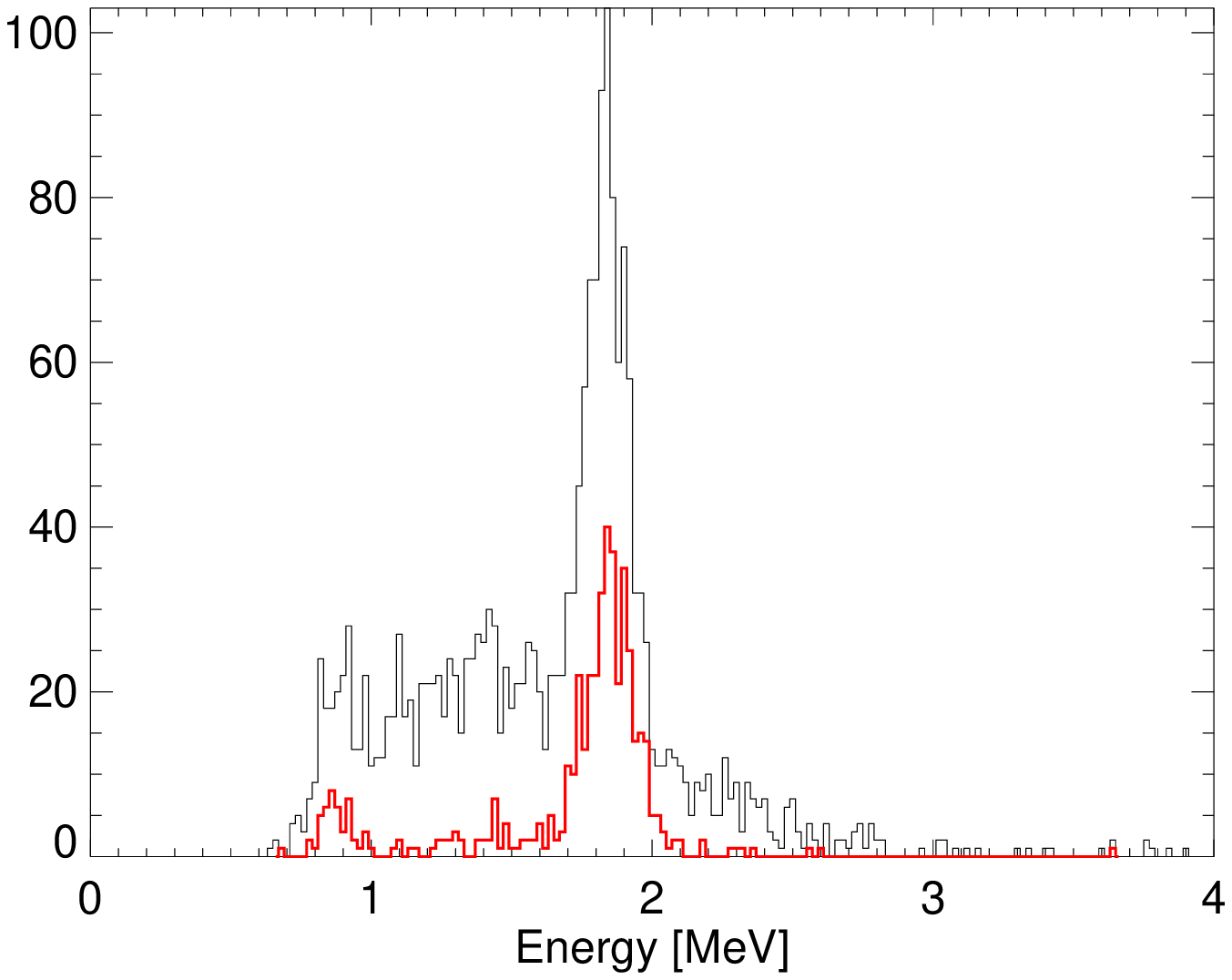,width=0.65\linewidth,
       bbllx=100,bblly=420,bburx=528,bbury=740,clip=}
\end{center}
\caption{\label{f:3stespenose} Energy spectrum for an \yt \ source at distance,
before any selection: the energy spectrum for events in the ARM peak is also
shown ({\it thick line}).} 
\efg
The CSR algorithm has been tested with a sample of 3--interaction events from an
\yt \ source (898 and 1836~\keV \ \g --rays) placed 2~m above the detector.
The energy spectrum for all the events in the 3-interaction sample, before
applying any selection, is shown in Fig.~\ref{f:3stespenose}.
Since the minimum energy threshold in this sample is $\sim$220~\keV \ (see
Fig.~\ref{f:3stespe}), the 898~\keV \ line turns out to be largely
suppressed. We note here that the minimum detectable energy is determined not
only by the detector {\it signal-to-noise} conditions, but also by the trigger
conditions and event selections applied during data taking. 
In Fig.~\ref{f:3stesep} the spatial separation between interactions is shown,
and in Fig.~\ref{f:3stez} the corresponding interaction depth. The distribution
for the first interaction peaks at high Z (i.e. small interaction depth), while
for the second interaction it peaks in the center and it is relatively flat for
the third interaction, with the maximum towards the bottom of the detector. This
is indeed what is expected for a source on the top of the detector, 2~m away.
\bfg[htp]
\begin{center}
\psfig{file=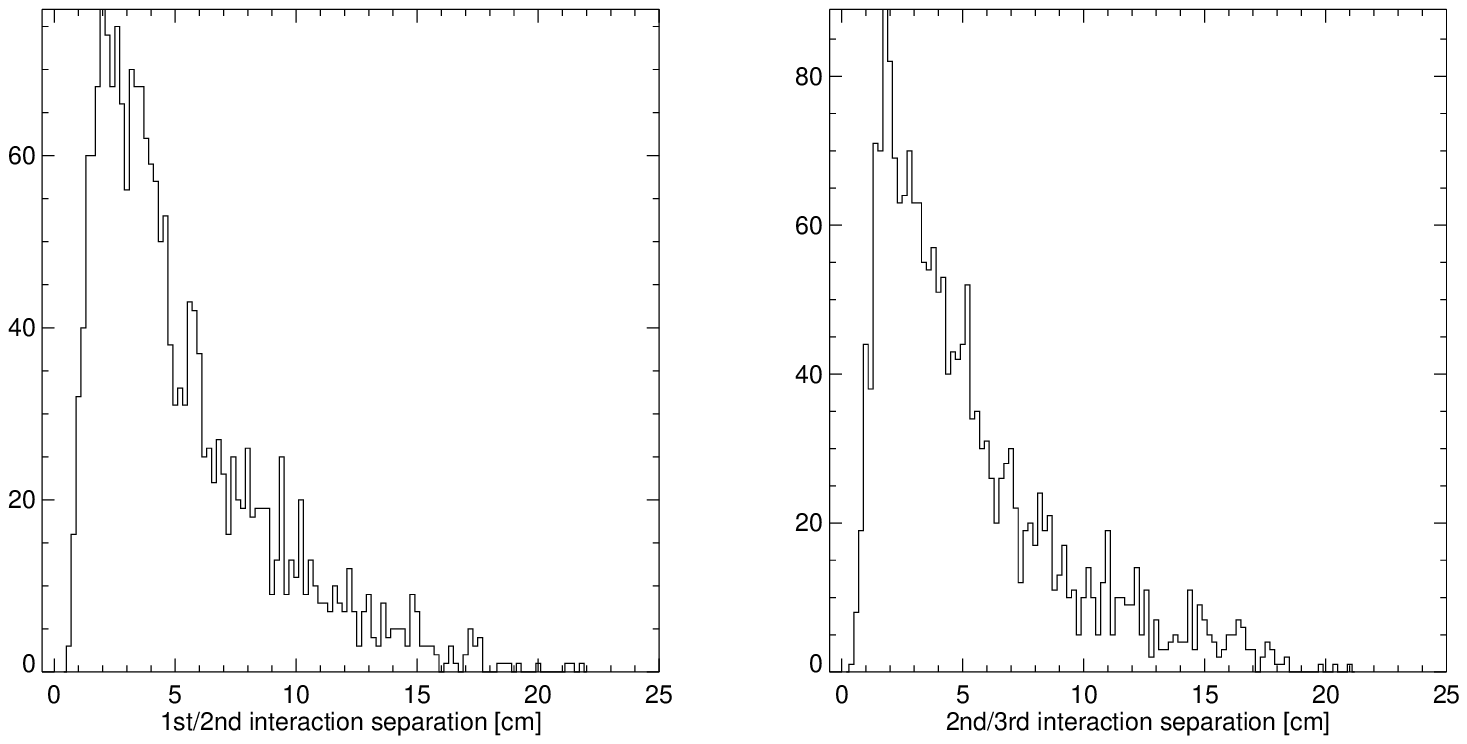,width=0.65\linewidth,height=5cm,
       bbllx=90,bblly=530,bburx=528,bbury=740,clip=}
\end{center}
\caption{\label{f:3stesep} Separation in space between interaction locations. {\it Left:}
first and second interaction, {\it right:} second and third interaction. The
3-interaction sequences were ordered by the CSR algorithm.}
\efg
\bfg[htp]
\begin{center}
\psfig{file=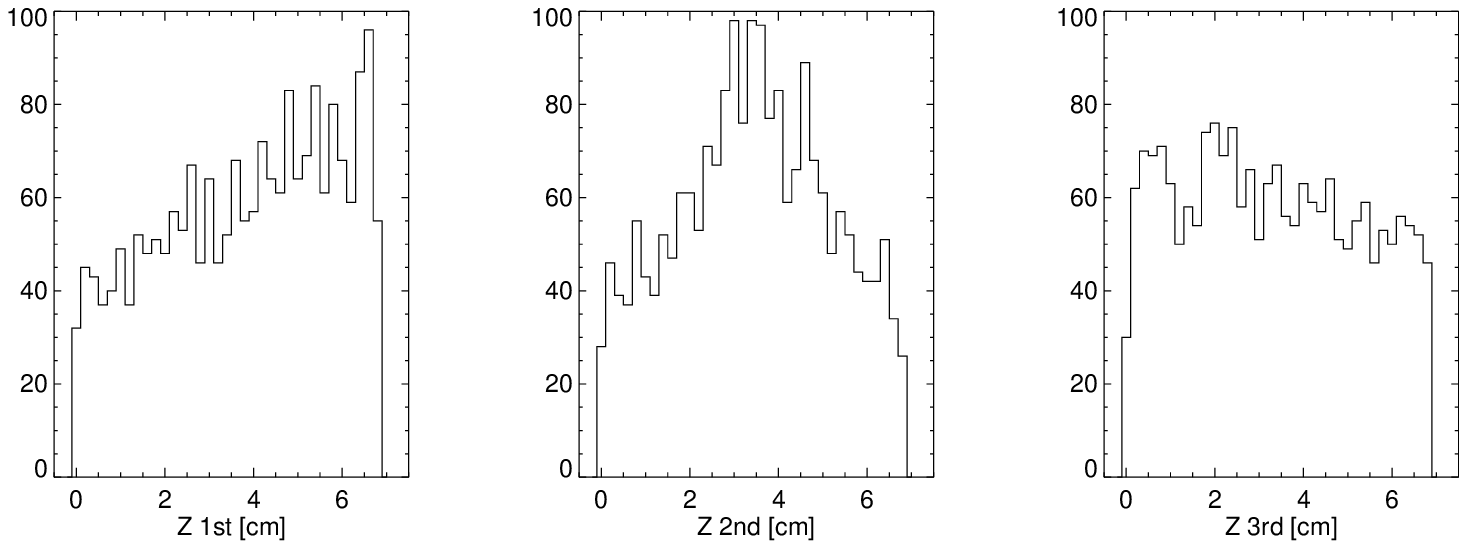,width=0.8\linewidth,
       bbllx=90,bblly=585,bburx=528,bbury=750,clip=}
\end{center}
\caption{\label{f:3stez} From left to right: interaction depth for the first,
second and third interaction location. The source was placed 2~m above the LXeTPC.
7~cm is the maximum drift length.}
\efg
\bfg[htb]
\begin{center}
\psfig{file=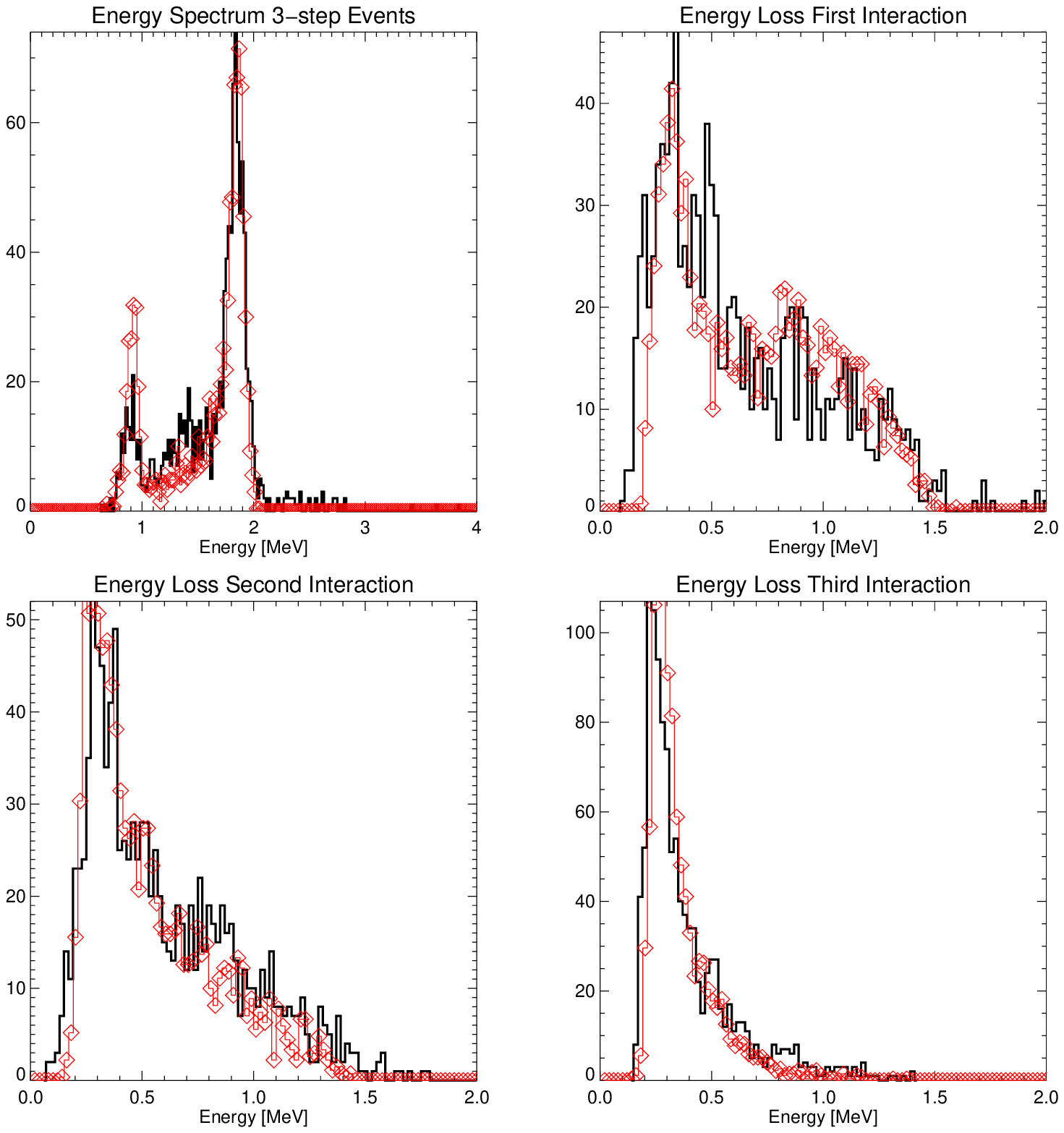,width=0.85\linewidth,
       bbllx=40,bblly=300,bburx=528,bbury=760,clip=}
\end{center}
\caption{\label{f:3stespe} 3-interaction events energy spectra for an \yt source
at distance (data selections as described in the text): solid thick line
$\rightarrow$ experimental data, diamonds $\rightarrow$ Monte Carlo data. 
{\it Left top:} energy spectrum as obtained summing up, for each event, the
energies in the 3 detected interactions ({\it sum-coincidence mode}). {\it Right
top:} energy spectrum for the first interaction. {\it Left bottom: } energy
spectrum for the second interaction. {\it Right bottom: } energy spectrum for
the third interaction.} 
\efg
\bfg[htb]
\begin{center}
\psfig{file=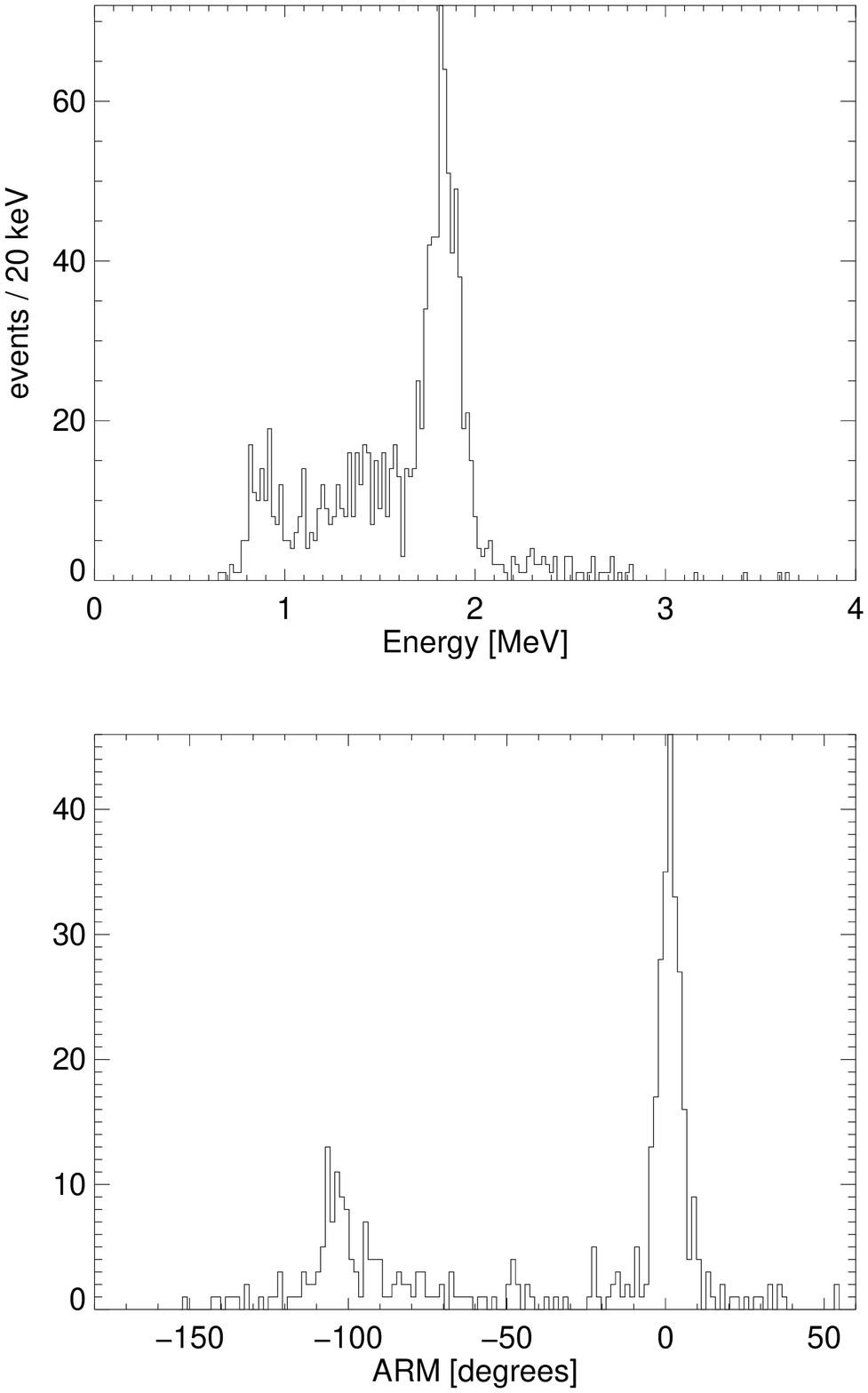,width=0.65\linewidth,
       bbllx=100,bblly=80,bburx=528,bbury=400,clip=}
\end{center}
\caption{\label{f:ARM} ARM spectrum for 1836~\keV \ photons (3-interaction
events): the interaction sequence has been ordered by the CSR algorithm
described in the paper. 55$\%$ of the events ends up in the ARM peak within
3$\sigma$'s from zero ($\sim$3 degrees).}
\efg
\begin{figure}[htb]
\bc
\epsfig{file=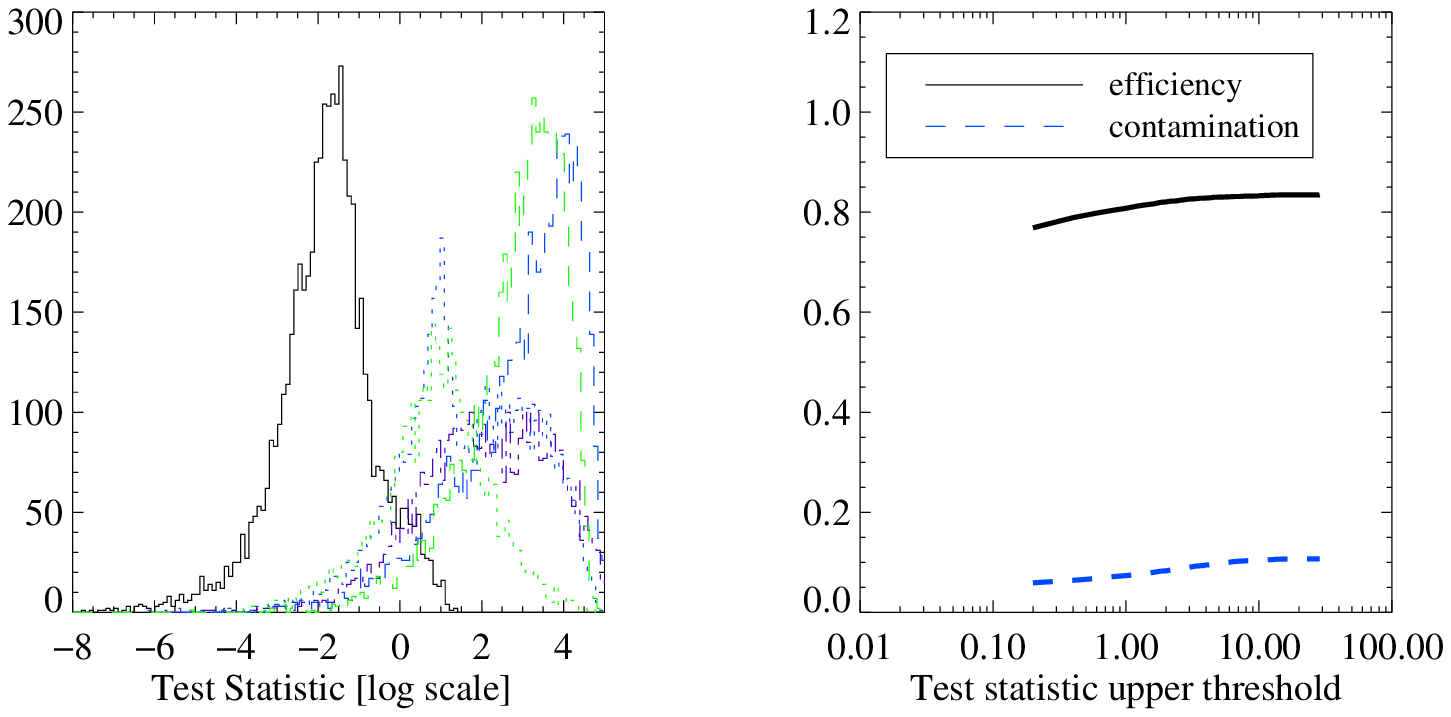,bbllx=100,bblly=530,bburx=540,bbury=750,width=0.9\linewidth,clip=}
\ec
\caption{\label{f:ACT} Monte Carlo generated events in the 2000~\keV \
photopeak, for the Xe-ACT described in the text: {\it left: } the test statistic
for the true sequence (continuos line) and for the remaining five false
sequences, {\it right: } fraction of correctly and incorrectly reconstructed
events ({\it efficiency} and {\it confusion}) as a function of upper threshold
on test statistic.} 
\end{figure}
Exploiting the
imaging capability of the detector, we impose a maximum separation of 7~cm
between interactions, in order to reject background events. Since the
attenuation length in LXe for 1~\MeV \ photon is 
$\sim$5~cm, separations larger than 7~cm are easily due to pile-up of
independent \g --rays. The impact of this selection can be evaluated comparing
the energy spectrum shown in Fig. \ref{f:3stespenose} to the one in
Fig. \ref{f:3stespe}, top-left panel, where events with energies exceeding
2~\MeV \ have been effectively rejected, as well as events in the Compton
continuum, and the 898~\keV \ line, even with reduced statistics, is clearly
detected. \\
Dealing with experimental data the true Compton sequence is not known {\it a
priori}, so that, to estimate the performance of the CSR algorithm, we have
necessarly to rely on additional information.\\
As a possible approach to test the performance of the CSR algorithm, we
compared experimental to Monte Carlo data (in this 
case 220~\keV \ energy threshold has been assumed also in the simulation):
Fig. \ref{f:3stespe} separately shows the distribution of the energies deposited
in the three interactions, as ordered in a sequence by the CSR algorithm, for
the case of \yt \ experimental
data and for Monte Carlo data for which the true sequence is known. The overall
agreement is fairly good: noticeably the three different spectra show
significantly distinguishable shape, making the comparison more sensitive. \\
A more direct approach to test the performance of the algorithm on experimental
data is the following: since the source position is
known, two indipendent measurements of the first Compton scattering angle are
given (\phibar$_1$ and \phigeo$_1$), translating into the so-called angular
resolution measure (ARM), defined as the difference between \phibar$_1$ and
\phigeo$_1$. Fig. \ref{f:ARM} shows the resulting ARM distribution obtained with
the 3--interaction events in the 1836~\keV \ full energy peak, ordered by the
CSR algorithm. For each event within $3~\sigma$s of the ARM peak around 0\deg,
$\sigma~\simeq~$3\deg, we assumed the sequence to have been correctly ordered,
while larger ARM values are considered wrongly reconstructed sequences. \\
Imposing an upper threshold of 2 
on the test statistic, we obtain 55$\%$ efficiency, in excellent 
agreement with the value expected from Monte Carlo. \\
The energy spectrum of these correctly sequenced events is shown in
Fig. \ref{f:3stespenose}: the Compton continuum and other
background events are clearly suppressed.

\section{From LXeGRIT to Xe-ACT}

A concept based on gas XeTPCs
has been recently proposed \cite{Rin} as a possible
candidate for an Advanced Compton Telescope (Xe-ACT), aiming at orders of magnitude
improvement over the current sensitivity, and further studies are
underway. Going from LXe to high-pressure Xe, a highly improved energy
resolution is obtained  \cite{ABolotnikov:97:hpxe}, approaching the
theoretical statistical limit. 20~\keV \ FWHM at 1~\MeV \ energy resolution has
been conservatively assumed, based on experimental data. Furthermore, in a lower
density active medium (e.g. 0.15~\gcc \ gas 
{\it vs.} 3.06~\gcc \ for LXe) the average separation between two consecutive
interactions is much larger, thus improving angular resolution. \\ 
A 2$\times$2$\times$2~m$^3$ active volume detector, filled with 0.15~\gcc \ Xe,
has been simulated. 
The performance of the CSR algorithm for 2~\MeV \ photons is shown in
Fig. \ref{f:ACT}: an efficiency exceeding 80$\%$ is achieved, with a
contamination level~$<~10\%$.

\section{Conclusions}

The efficiency of the LXeGRIT instrument as a Compton telescope depends on the
capability to order the sequence of multiple Compton
interactions detected in the TPC. This capability has been experimentally
tested, and we conclude that the true 
sequence can be efficiently (efficiency $\sim 60 \%$ for 1.8~\MeV \ photons)
determined, in good agreement with expectations (i.e. Monte Carlo model). The
experience gained in operating the LXeGRIT detector also suggests possible ways
to improve this performance based on realistic detector conditions.  

\acknowledgments  

We would like to thank Ian Mulvany and Burair Kothari of Columbia University
for their contributions to the implementation of the Compton sequence
reconstruction algorithm.\\ 
This work was supported by NASA under grant NAG5-5108.

\small
\bibliography{CSR_last}  
\bibliographystyle{spiebib}   

\end{document}